\documentclass[preprint, superscriptaddress, showpacs,preprintnumbers,amsmath,amssymb,prd]{revtex4}
\usepackage{graphicx}

\begin{document}

\thispagestyle{empty}

\title{Stronger constraints on axion from measuring the Casimir
interaction by means of dynamic atomic force microscope
}

\author{V.~B.~Bezerra}
\affiliation{Department of Physics, Federal University of Para\'{\i}ba,
C.P.5008, CEP 58059--970, Jo\~{a}o Pessoa, Pb-Brazil}
\author{
G.~L.~Klimchitskaya}
\affiliation{Central Astronomical Observatory
at Pulkovo of the Russian Academy of Sciences,
St.Petersburg, 196140, Russia}
\affiliation{Institute of Physics, Nanotechnology and
Telecommunications, St.Petersburg State
Polytechnical University, St.Petersburg, 195251, Russia}
\affiliation{Department of Physics, Federal University of Para\'{\i}ba,
C.P.5008, CEP 58059--970, Jo\~{a}o Pessoa, Pb-Brazil}
\author{
 V.~M.~Mostepanenko}
\affiliation{Central Astronomical Observatory
at Pulkovo of the Russian Academy of Sciences,
St.Petersburg, 196140, Russia}
\affiliation{Institute of Physics, Nanotechnology and
Telecommunications, St.Petersburg State
Polytechnical University, St.Petersburg, 195251, Russia}
\affiliation{Department of Physics, Federal University of Para\'{\i}ba,
C.P.5008, CEP 58059--970, Jo\~{a}o Pessoa, Pb-Brazil}
\author{C.~Romero}
\affiliation{Department of Physics, Federal University of Para\'{\i}ba,
C.P.5008, CEP 58059--970, Jo\~{a}o Pessoa, Pb-Brazil}

\begin{abstract}
We calculate the additional force due to two-axion exchange
acting in a sphere-disc geometry, used in experiments on measuring
the gradient of the Casimir force. With this result, stronger
constraints on the pseudoscalar coupling constants of an axion
and axion-like particles to
a proton and a neutron are obtained over the wide range of
axion masses from $3\times 10^{-5}$ to 1\,eV.
Among the three experiments with Au-Au, Au-Ni and Ni-Ni boundary
surfaces performed by means of dynamic atomic force microscope,
major improving is achieved for the experiment with Au-Au test
bodies. Here, the constraints obtained are stronger up to a
factor of 170, as compared to the previously known ones.
The largest strengthening holds for the axion mass 0.3\,eV.
\end{abstract}
\pacs{14.80.Va, 12.20.Fv, 14.80.-j}

\maketitle
\section{Introduction}

It happens that some predicted particle is of
primary importance for theoretical reasons, but avoids experimental
observation over many years. Commonly known predictions of
this kind, which were finally confirmed, are the neutrino and
the Higgs boson. In a similar way, the axion was predicted
\cite{1,2}
in 1978 as a consequence of breaking of Peccei and Quinn
symmetry \cite{3}, but has not been discovered up to the present.
It is known that  Peccei and Quinn symmetry provides
the most natural possibility to avoid strong {\it CP}
violation and the
large electric dipole moment induced by it for the neutron in quantum
chromodynamics.
What is more, axion reasonably explains the nature of dark matter
in astrophysics and cosmology \cite{3a,4o}.
Because of this, after the  proper QCD axions
were constrained to a narrow band in parameter space,
 different axion-like
fields have been introduced, specifically, inspired by string
theory \cite{4pp}.
Some of them can be used to solve the problem of strong
{\it CP} violation (see, for instance, the model of GUT axion
\cite{4a,4b} or various variants \cite{4c,4d} of the model
of hadronic axion \cite{4e,4f}),
 but not all. Keeping in mind, however, that our
approach in the search for axion and axion-like particles is
common, below we use both terms more or less synonymously.
There is an opinion that
eventually the existence of the axion will be confirmed.
For this reason, many experiments searching for the axion
have been performed in the past and many more are planed for
the near future.

It has been suggested \cite{4,5,6} to use the gravitational
experiments of E\"{o}tvos \cite{6,7} and Cavendish \cite{8}
type for constraining the coupling constants of the
pseudoscalar interaction of massless axions to protons and
neutrons. This approach was generalized \cite{9} for the more
realistic case of massive axions, and rather strong constraints
were obtained in the mass range $m_a\leq 9.9\,\mu$eV.
Constraints on the coupling constants of an axion to
nucleons were also obtained from searching the Compton-like
process for axions emitted from the Sun \cite{10} and from
observations of the neutrino burst of supernova 1987A \cite{11}.
Cosmology also places constraints on axion-like
particles in different ranges
of their mass if we take into account that axions could
have been produced
at the early stages of Universe evolution and contribute to the
dark matter. Thus, thermal axions with $m_a>0.7\,$eV are
excluded because they would make too large contribution to the
hot dark matter \cite{12}
with a reservation that such kind limits typically refer to
specific couplings in some axion models and may be not applicable
to all axion-like particles.
At the same time, some models of
cold dark matter exclude axions with $m_a<10\,\mu$eV \cite{10}.

The interaction with photons is used in many experimental searches
for axions. Thus, strong constraints on the coupling constant of an
axion to photon were obtained from searching axions emitted from
the Sun by means of the CERN axion solar telescope \cite{13}.
These constraints are relevant to axions with masses from
$m_a=0.39\,$eV to $m_a=0.64\,$eV.
Strong astrophysical constraints on the axion-photon interaction
were found from gravitationally bound systems of stars of
approximately the same age (globular clusters).
These constraints are valid for axions with
larger masses from about 0.3\,eV to several tens keV \cite{10,14}
(too heavy axions  are also excluded due to
their enormously large contribution to hot dark matter).

Besides nucleons and photons, axion may also interact with electrons.
Interaction with electrons includes the Compton process and the
electron-positron annihilation with emission of an axion in stellar
plasmas. These processes are used for constraining the
axion-electron coupling constant \cite{15,16}.
As was mentioned above, experiments on the search of axions are
numerous and varied (see reviews in Refs.~\cite{4pp,17,18,19}).
However, there is still the so-called {\it window} in the possible
values of axion masses which extends \cite{10} from approximately
$m_a=10^{-5}\,$eV to $m_a=10^{-2}\,$eV, where constraints on the
axion parameters are rather weak (haloscope searches aim to
narrow this window \cite{4pp}).

In Ref.~\cite{20} strong constraints on the coupling constants of
pseudoscalar interaction of axion to nucleons in the mass range
from $10^{-4}\,$eV to 0.3\,eV were obtained from measurements of
the thermal Casimir-Polder force between a Bose-Einstein
condensate of ${}^{87}$Rb atoms and a SiO${}_2$ plate \cite{21}.
Experiments on measuring the Casimir force \cite{22} have long
been used for obtaining constraints on the Yukawa-type corrections
to Newton's gravitational law
(see review in Ref.~\cite{23} and Refs.~\cite{24,25,26,27aa,27,28aa}
for the most recent results). The Yukawa-type correction terms to
Newtonian gravity arise either due to exchange of light scalar
particles between atoms of test bodies or  from extra-dimensional
physics with low-energy compactification scale \cite{27a}.
The axion is a
pseudoscalar particle. As a result, the effective potential between
two fermions, arising from the exchange of an axion with the
pseudoscalar coupling to fermions, is spin-dependent \cite{9}.
Keeping in mind that the test bodies in Casimir experiments are
unpolarized, one concludes that there is no any extra force due to
a single-axion exchange. In Ref.~\cite{20} the additional
spin-independent force between an atom and a plate was found as
a result of two-axion exchange (in Ref.~\cite{9} this process
was used for constraining the parameters of an axion from the
gravitational experiments of E\"{o}tvos- and Cavendish-type).

In this paper, we obtain stronger constraints on the constants of
pseudoscalar coupling of axion-like particles
to a proton and a neutron which
follow from experiments on measuring the gradient of the Casimir
force between a sphere and a plate by means of dynamic atomic force
microscope (AFM). We derive constraints following from the
experiments with an Au-coated sphere and an Au-coated plate
\cite{28,28a}, with an Au-coated sphere and a Ni-coated plate
\cite{29}, and, finally, from the experiment with two Ni-coated
test bodies \cite{30,31}. For this purpose, we calculate the
gradient of the additional force arising between a sphere and
a plate due to two-axion exchange taking into account the layer
structure of both test bodies used in each experiment.
The constraints obtained from each of the three experiments in the
region of axion masses from $m_a=2\times 10^{-5}\,$eV to
$m_a=1\,$eV are compared between themselves and with those
of Ref.~\cite{20}. It is shown that in the region of axion
masses from $m_a=10^{-4}\,$eV to $m_a=0.3\,$eV, which was
covered in Ref.~\cite{20}, the constraints obtained
in this work are
significantly stronger. The largest strengthening by a factor
of 170 holds for an axion mass $m_a=0.3\,$eV.

The paper is organized as follows. In Sec.~II, we derive
an expression for the gradient of the force arising between
a sphere and a plate due to two-axion exchange.
In Sec.~III, this expression is adapted for the layer
structure of the experiment with two Au-coated test bodies,
and the respective constraints on the axion-nucleon coupling
constants are obtained. In Sec.~IV, the same is done for the
experiment with an Au-coated sphere and a Ni-coated plate.
Section~V contains derivation of the constraints following from
the experiment with two Ni-coated test bodies. In Sec.~VI
the reader will find our conclusions and discussion.

In what follows, the system of units with $\hbar=c=1$ is used.

\section{Gradient of the force between a sphere and a plate
due to two-axion exchange}

In three experiments performed by means of AFM
\cite{28,28a,29,30,31} the gradient of the Casimir force was
measured between spheres of about $50\,\mu$m radius and discs
of approximately 5\,mm radius made of different materials.
Taking into account that the characteristic size of the
sphere is by a factor of 100 smaller than that of the disc,
all points of the sphere can be considered as situated
above the disc center (i.e., the disc area can be considered
as infinitely large). Here, we calculate the additional force
acting in a sphere-plate configuration due to two-axion
exchange between protons and neutrons belonging to these test
bodies. In doing so, we assume the pseudoscalar character of
an axion-fermion interaction and neglect the interaction
of axions with electrons \cite{9}. In any case, the account
of possible scalar coupling of axions to fermions \cite{32}
or of axion-electron interaction
could only slightly increase the force
magnitude and, as a result, would lead to minor strengthening
of the constraints obtained (see Refs.~\cite{33,34} for the
constraints on coupling constants of scalar interaction of
axion with nucleons).

Let the coordinate plane $x,\,y$ coincide with an upper plane
surface of the disc of thickness $D$ and let the $z$ axis be
perpendicular to it. We choose the origin of the coordinate
system at the center of the upper surface of the disc.
The sphere of radius $R$ has its center at $z=a+R$.
The effective potential due to two-axion exchange between
two nucleons (protons or neutrons) situated at the points
$\mbox{\boldmath$r$}_2$ of the disc and
$\mbox{\boldmath$r$}_1$ of the sphere is given by
\cite{9,35,36}
\begin{equation}
V(|\mbox{\boldmath$r$}_1-\mbox{\boldmath$r$}_2|)=
-\frac{g_{ak}^2g_{al}^2}{32\pi^3m^2}\,
\frac{m_a}{(\mbox{\boldmath$r$}_1-\mbox{\boldmath$r$}_2)^2}
K_1(2m_a|\mbox{\boldmath$r$}_1-\mbox{\boldmath$r$}_2|).
\label{eq1}
\end{equation}
\noindent
Here, $g_{ak}$ and $g_{al}$ are the constants of a pseudoscalar
axion-proton ($k,\,l=p$) or axion-neutron ($k,\,l=n$)
interaction, $m=(m_n+m_p)/2$ is the mean of the neutron and
proton masses, and $K_1(z)$ is the modified Bessel function
of the second kind. In the derivation of Eq.~(\ref{eq1}) it was
assumed that
$|\mbox{\boldmath$r$}_1-\mbox{\boldmath$r$}_2|\gg 1/m$.
This condition is fulfiled with large safety margin because
in the experiments of Refs.~\cite{28,28a,29,30,31}
it holds $a>200\,$nm.

Using Eq.~(\ref{eq1}), the additional force due to
two-axion exchange, $F_{\rm add}$, can be obtained by integration
over the volumes of the sphere $V_s$ and of the disc $V_d$
\begin{equation}
F_{\rm add}(a)=-\int_{V_s}\!d\mbox{\boldmath$r$}_1
\int_{V_d}\!d\mbox{\boldmath$r$}_2
\frac{\partial}{\partial z_1}
V(|\mbox{\boldmath$r$}_1-\mbox{\boldmath$r$}_2|).
\label{eq2}
\end{equation}

Now we introduce the polar coordinates on the disc and calculate
the gradient of the force (\ref{eq2}) taking
into account Eq.~(\ref{eq1})
\begin{eqnarray}
&&
\frac{\partial F_{\rm add}(a)}{\partial a}=
\frac{\pi m_a}{m^2m_H^2}C_dC_s\frac{\partial}{\partial a}
\int_{a}^{2R+a}\!\!dz_1[R^2-(z_1-R-a)^2]
\nonumber \\
&&~~~
\times\frac{\partial}{\partial z_1}\int_{-D}^{0}\!dz_2
\int_{0}^{\infty}\!\!\!\rho d\rho
\frac{K_1(2m_a\sqrt{\rho^2+(z_1-z_2)^2})}{\rho^2+(z_1-z_2)^2}\,.
\label{eq3}
\end{eqnarray}
\noindent
Here, the coefficient $C_d$ for the disc material is defined as
\begin{equation}
C_d=\rho_d\left(\frac{g_{ap}^2}{4\pi}\,
\frac{Z_d}{\mu_d}+\frac{g_{an}^2}{4\pi}\,
\frac{N_d}{\mu_d}\right),
\label{eq4}
\end{equation}
\noindent
where $\rho_d$ is the disc density, and $Z_d$ and $N_d$ are
the number of protons and the mean number of neutrons in the
disc atom (molecule). The quantity $\mu_d=m_d/m_H$, where
$m_d$ and $m_H$ are the mean  mass of the disc atom (molecule)
and the mass of an atomic hydrogen, respectively.
Note that the values of $Z/\mu$ and $N/\mu$ for the first 92
elements of the Periodic Table with account of their isotopic
composition are presented in Ref.~\cite{37}. This information
is used below in computations of Secs.~III--V.
The coefficient $C_s$ for a sphere material is defined similarly
to Eq.~(\ref{eq4}) with a replacement of all indices $d$ for $s$.

It is convenient to calculate first the quantity
\begin{equation}
I\equiv\frac{\partial}{\partial z_1}\int_{-D}^{0}\!dz_2
\int_{0}^{\infty}\!\!\!\rho d\rho
\frac{K_1(2m_a\sqrt{\rho^2+(z_1-z_2)^2})}{\rho^2+(z_1-z_2)^2}
\label{eq5}
\end{equation}
\noindent
entering Eq.~(\ref{eq3}). Using the integral representation
\cite{38}
\begin{equation}
\frac{K_1(z)}{z}=\int_{1}^{\infty}\!\!\!du\sqrt{u^2-1}e^{-zu},
\label{eq6}
\end{equation}
\noindent
one can rearrange Eq.~(\ref{eq5}) in the form
\begin{equation}
I=2m_a\int_{1}^{\infty}\!\!\!du\sqrt{u^2-1}
\frac{\partial}{\partial z_1}\int_{-D}^{0}\!\!dz_2
\int_{0}^{\infty}\!\!\!\!\rho d\rho
\frac{e^{-2m_au\sqrt{\rho^2+(z_1-z_2)^2}}}{\sqrt{\rho^2+(z_1-z_2)^2}}.
\label{eq7}
\end{equation}

Now we introduce the new variable $v=\sqrt{\rho^2+(z_1-z_2)^2}$
instead of $\rho$ in Eq.~(\ref{eq7}) and integrate with respect to
it
\begin{equation}
I=\int_{1}^{\infty}\!du\frac{\sqrt{u^2-1}}{u}
\frac{\partial}{\partial z_1}\int_{-D}^{0}\!\!dz_2
e^{-2m_au(z_1-z_2)}.
\label{eq8}
\end{equation}
\noindent
Integrating and differentiating  in Eq.~(\ref{eq8}) with respect
to $z_2$ and $z_1$, respectively, one obtains
\begin{equation}
I=-\int_{1}^{\infty}\!du\frac{\sqrt{u^2-1}}{u}
e^{-2m_auz_1}\left(1-e^{-2m_auD}\right).
\label{eq9}
\end{equation}

Substituting Eq.~(\ref{eq9}) in Eq.~(\ref{eq3}) and
differentiating
with respect to $a$, we arrive at
\begin{eqnarray}
&&
\frac{\partial F_{\rm add}(a)}{\partial a}=
\frac{2\pi m_a}{m^2m_H^2}C_dC_s
\int_{1}^{\infty}\!du\frac{\sqrt{u^2-1}}{u}
\left(1-e^{-2m_auD}\right)
\nonumber \\
&&~~~~~~~
\times\int_{a}^{2R+a}\!\!dz_1(R+a-z_1)e^{-2m_auz_1}.
\label{eq10}
\end{eqnarray}
\noindent
Finally, calculating the integral with respect to $z_1$,
we find
\begin{eqnarray}
&&
\frac{\partial F_{\rm add}(a)}{\partial a}=
\frac{\pi}{m^2m_H^2}C_dC_s
\int_{1}^{\infty}\!du\frac{\sqrt{u^2-1}}{u^2}
\left(1-e^{-2m_auD}\right)
\nonumber \\
&&~~~~~~~
\times
e^{-2m_aau}\,\Phi(R,m_au),
\label{eq11}
\end{eqnarray}
\noindent
where the following notation is introduced
\begin{equation}
\Phi(r,z)=r-\frac{1}{2z}+e^{-2rz}\left(r+
\frac{1}{2z}\right).
\label{eq12}
\end{equation}

{}From Eq.~(\ref{eq11}) one can simply obtain the additional
force gradient due to two-axion exchange between a spherical
envelope of radius $R$ and thickness $\Delta_s$ and a plate.
This can be found by subtracting from Eq.~(\ref{eq11}) the
force due to a sphere of radius $R-\Delta_s$ placed at a
separation $a+\Delta_s$ from the plate:
\begin{eqnarray}
&&
\frac{\partial F_{\rm add}(a)}{\partial a}=
\frac{\pi}{m^2m_H^2}C_dC_s
\int_{1}^{\infty}\!du\frac{\sqrt{u^2-1}}{u^2}
\left(1-e^{-2m_auD}\right)
\nonumber \\
&&~~~~
\times
e^{-2m_aau}\,\left[\Phi(R,m_au)-
e^{-2m_au\Delta_s}\Phi(R-\Delta_s,m_au)\right].
\label{eq13}
\end{eqnarray}

In the end of this section, we note that the relative
contribution of the boundary effects arising due to a
finiteness of the plate area to the force gradient
(\ref{eq13}) is of the order
\begin{equation}
\delta\left(\frac{\partial F_{\rm add}}{\partial a}
\right)\sim\frac{R}{R_d^2m_a},
\label{eq14}
\end{equation}
\noindent
where $R_d$ is the disc radius. Taking into account the values
of the experimental parameters indicated above, one can conclude
that the contribution of the boundary effects in the additional
force gradient  (\ref{eq13}) does not exceed 1\%
for axion masses satisfying the condition
$m_a>1/R_d\approx 4\times 10^{-5}\,$eV.

\section{Constraints from experiment with two
gold-coated
test bodies}

In the experiment of Refs.~\cite{28,28a}, the dynamic AFM was
used to measure the gradient of the Casimir force between an
Au-coated hollow sphere (spherical envelope) of
$R=41.3\,\mu$m radius made of fused silica and an Au-coated
thick sapphire disc of $R_d=5\,$mm radius.
The thickness of glass spherical envelope was
$\Delta_s^{\!g}=5\,\mu$m.
The thicknesses and density
of the Au coatings on the sphere and the disc were
$\Delta_s^{\!\rm Au}=\Delta_d^{\!\rm Au}=280\,$nm and
$\rho_{\rm Au}=19.29\,\mbox{g/cm}^3$, respectively.
In relation to the Casimir force this thickness is quite
sufficient for considering the sphere and the disc as
all-gold bodies \cite{23}.
However, the additional force due to two-axion exchange is
highly sensitive to the material composition of both test bodies,
and this should be taken into account in computations.
In the experiment of Refs.~\cite{28,28a}, the Au layers on
both test bodies were deposited on the layer of Al of thickness
$\Delta_s^{\!\rm Al}=\Delta_d^{\!\rm Al}=20\,$nm with
density $\rho_{\rm Al}=2.7\,\mbox{g/cm}^3$.
The density of glass (SiO${}_2$, the material of the spherical
envelope) was $\rho_{g}=2.5\,\mbox{g/cm}^3$, whereas the
density of sapphire (Al${}_2$O${}_3$, the material of the disc)
was $\rho_{\rm sa}=4.1\,\mbox{g/cm}^3$.

By applying Eq.~(\ref{eq13}) to each pair of material layers
forming the spherical envelope and the disc, we arrive at the
following expression for the gradient of the additional force
arising from the two-axion exchange:
\begin{eqnarray}
&&
\frac{\partial F_{\rm add}(a)}{\partial a}=
\frac{\pi}{m^2m_H^2}
\int_{1}^{\infty}\!du\frac{\sqrt{u^2-1}}{u^2}
e^{-2m_aau}
\nonumber \\
&&~~~~~~~~~~
\times
X_s(m_au)X_d(m_au),
\label{eq15}
\end{eqnarray}
\noindent
where the functions $X_d$ and $X_s$ are defined as follows:
\begin{eqnarray}
&&
X_d(z)\equiv C_{\rm Au}\left(1-e^{-2z\Delta_d^{\!\rm Au}}\right)
+C_{\rm Al}e^{-2z\Delta_d^{\!\rm Au}}\left(1-e^{-2z\Delta_d^{\!\rm Al}}\right)
+C_{\rm sa}e^{-2z(\Delta_d^{\!\rm Au}+\Delta_d^{\!\rm Al})},
\nonumber \\[-1mm]
&&
\label{eq16} \\[-2mm]
&&
X_s(z)\equiv C_{\rm Au}\left[\Phi(R,z)-e^{-2z\Delta_s^{\!\rm Au}}
\Phi(R-\Delta_s^{\!\rm Au},z)\right]
\nonumber \\
&&~~~
+C_{\rm Al}e^{-2z\Delta_s^{\!\rm Au}}
\left[\Phi(R-\Delta_s^{\!\rm Au},z)-e^{-2z\Delta_s^{\!\rm Al}}
\Phi(R-\Delta_s^{\!\rm Au}-\Delta_s^{\!\rm Al},z)\right]
\nonumber \\
&&~~~
+C_{g}e^{-2z(\Delta_s^{\!\rm Au}+\Delta_s^{\!\rm Al})}
\left[\Phi(R-\Delta_s^{\!\rm Au}-\Delta_s^{\!\rm Al},z)-
e^{-2z\Delta_s^{\! g}}
\Phi(R-\Delta_s^{\!\rm Au}-\Delta_s^{\!\rm Al}
-\Delta_s^{\!g},z)\right].
\nonumber
\end{eqnarray}
\noindent
Note that the thickness of the sapphire disc is put equal to
infinity, as it does not influence the result.
The coefficients $C_{\rm Au}$,  $C_{\rm Al}$,  $C_{\rm sa}$
and  $C_{g}$ are defined in Eq.~(\ref{eq4}), as applied to the
atoms Au, Al and to the molecules Al${}_2$O${}_3$, SiO${}_2$,
respectively. In Ref.~\cite{37} one finds the values of
the following multiples entering Eq.~(\ref{eq4}):
\begin{eqnarray}
&&
\frac{Z_{\rm Au}}{\mu_{\rm Au}}=0.40422,\qquad
\frac{N_{\rm Au}}{\mu_{\rm Au}}=0.60378,
\label{eq17} \\
&&
\frac{Z_{\rm Al}}{\mu_{\rm Al}}=0.48558,\qquad
\frac{N_{\rm Al}}{\mu_{\rm Al}}=0.52304.
\nonumber
\end{eqnarray}
\noindent
For a sapphire molecule Al${}_2$O${}_3$, in
accordance to Ref.~\cite{37}, we have
\begin{eqnarray}
&&
\frac{Z_{\rm Al_2O_3}}{\mu_{\rm Al_2O_3}}=
\frac{2Z_{\rm Al}+3Z_{\rm O}}{2\mu_{\rm Al}+
3\mu_{\rm O}}=0.49422,
\label{eq18} \\
&&
\frac{N_{\rm Al_2O_3}}{\mu_{\rm Al_2O_3}}=
\frac{2N_{\rm Al}+3N_{\rm O}}{2\mu_{\rm Al}+
3\mu_{\rm O}}=0.51412.
\nonumber
\end{eqnarray}
\noindent
In a similar way, for fused silica molecule SiO${}_2$
we obtain \cite{20,37}
\begin{equation}
\frac{Z_{\rm SiO_2}}{\mu_{\rm SiO_2}}=0.503205,\qquad
\frac{N_{\rm SiO_2}}{\mu_{\rm SiO_2}}=0.505179.
\label{eq19}
\end{equation}
\noindent

The experimental data of Refs.~\cite{28,28a} for the gradient of
the Casimir force $F_C^{\prime}(a)$ were measured at separation
distances $a\geq 235\,$nm and were found to be in agreement with
theoretical predictions of the Lifshitz theory of dispersion
forces \cite{23,39} under the condition
that the low-frequency behavior of the dielectric permittivity
of Au is described by the plasma model. This means that the
gradient of any additional force due to two-axion exchange is
constrained by the inequality
\begin{equation}
\frac{\partial F_{\rm add}(a)}{\partial a}\leq
\Delta F_C^{\prime}(a).
\label{eq20}
\end{equation}
\noindent
Here, $\Delta F_C^{\prime}(a)$ is the total absolute error
in the measured gradient of the Casimir force, which was
determined in Refs.~\cite{28,28a} as a combination of random
and systematic errors at a 67\% confidence level.

We have substituted Eq.~(\ref{eq15}) in Eq.~(\ref{eq20}) and
found numerically the regions of the axion parameters,
$g_{ap}$, $g_{an}$ and $m_a$, where the inequality (\ref{eq20})
is satisfied. The most strong constraints were obtained at the
shortest experimental separation $a=235\,$nm, where the
absolute error
$\Delta F_C^{\prime}=0.5\,\mu$N/m \cite{28,28a}.
Note that the gradient of the Casimir force at this separation
is equal to $F_C^{\prime}=73.86\,\mu$N/m.
In Fig.~1, we present the obtained constraints on the
pseudoscalar coupling constants $g_{ap(n)}^2/(4\pi)$
as a function of the axion mass $m_a$.
The solid lines from bottom to top are plotted under the
conditions $g_{ap}^2=g_{an}^2$, $g_{an}^2\gg g_{ap}^2$
and $g_{ap}^2\gg g_{an}^2$, respectively.
The regions above each line are prohibited by the
measurement data of Refs.~\cite{28,28a}, and the regions below each
line are allowed by the data. In Table~I, the maximum
allowed values of the pseudoscalar coupling constants of
an axion-nucleon interaction
for different values of the axion mass
(column 1) are listed under the condition
$g_{ap}^2=g_{an}^2$ (column 2),
$g_{an}^2\gg g_{ap}^2$ (column 3), and
$g_{ap}^2\gg g_{an}^2$ (column 4).
As can be seen in Fig.~1, for $m_a\leq 0.001\,$eV the obtained
constraints are almost independent of the axion mass
(we do not extend our constraints to smaller $m_a$ where the
omitted boundary effects may exceed 1\%).

In Fig.~2, we compare the constraints of Fig.~1 and Table~I,
found under the most reasonable condition $g_{ap}^2=g_{an}^2$,
with the constraints of Ref.~\cite{20} obtained from
measurements of the thermal Casimir-Polder force between the
Bose-Einstein condensate of ${}^{87}$Rb atoms and SiO${}_2$
plate \cite{21}. The solid line reproduces the lowest line
 of Fig.~1 within an interval
 $10^{-4}\,\mbox{eV}\leq m_a\leq 0.3\,$eV, where the
 constraints of Ref.~\cite{20} were obtained, and the dashed
 line shows the respective constraints of Ref.~\cite{20}.
 As can be seen in Fig.~2, the experiment on measuring the
 gradient of the Casimir force by means of dynamic AFM
 \cite{28,28a} leads to stronger constraints than the
 experiment of Ref.~\cite{21} over the entire mass
 range of the latter.
 The largest strengthening by a factor of 170 holds for the
 axion mass $m_a=0.3\,$eV. In the region of small axion
 masses $m_a\lesssim 0.003\,$eV, the strengthening is by a
 factor of 1.2. Similar results are obtained under
alternative assumptions about the relationship between
$g_{ap}$ and $g_{an}$. Thus, if $g_{an}\gg g_{ap}$,
the constraints of Fig.~1 are stronger than those of
Ref.~\cite{20} up to a factor 185.
The largest strengthening is achieved at the same
 axion mass $m_a=0.3\,$eV.
Under the assumption $g_{ap}\gg g_{an}$, the respective
strengthening is up to a factor 150 and again holds
at $m_a=0.3\,$eV.
These results demonstrate that measurements of the gradient
of the Casimir force by means of dynamic AFM should be
considered as a usefull supplementary method in the search
for an axion.

\section{Constraints from experiment with gold-coated
sphere and nickel-coated plate}

In Ref.~\cite{29}, the dynamic AFM was
used to measure the gradient of the Casimir force between an
Au-coated hollow sphere of
$R=64.1\,\mu$m radius made of fused silica and a Ni-coated
Si disc of density $\rho_{\rm Si}=2.33\,\mbox{g/cm}^3$,
which, again, can be
considered as infinitely large.
The thickness of glass spherical envelope was
$\Delta_s^{\!g}=5\,\mu$m, i.e., the same as in the experiment
of Refs.~\cite{28,28a}. It was coated by the layers of Al and
Au of the same thicknesses $\Delta_s^{\!\rm Al}$ and
$\Delta_s^{\!\rm Au}$, as in the experiment with two Au
surfaces. The Si disc was coated with a single layer of
magnetic metal Ni having the thickness
$\Delta_d^{\!\rm Ni}=154\,$nm and
density $\rho_{\rm Ni}=8.9\,\mbox{g/cm}^3$.

The gradient of the additional force between the hollow
sphere and the plate due to two-axion exchange in this
experiment is, again, given by Eq.~(\ref{eq15}), where the
function $X_s$ is presented in Eq.~(\ref{eq16}).
As to the function $X_d$, it takes a more simple form than
in Eq.~(\ref{eq16}), due to the presence of only one coating
layer on the disc surface. For this function one obtains
\begin{equation}
X_d(z)= C_{\rm Ni}\left(1-e^{-2z\Delta_d^{\!\rm Ni}}\right)
+C_{\rm Si}e^{-2z\Delta_d^{\!\rm Ni}}.
\label{eq21}
\end{equation}

The coefficients $C_{\rm Ni}$ and $C_{\rm Si}$ are defined by
Eq.~(\ref{eq4}) where the values $Z/\mu$ and $N/\mu$ are
given by \cite{37}
\begin{eqnarray}
&&
\frac{Z_{\rm Ni}}{\mu_{\rm Ni}}=0.48069,\qquad
\frac{N_{\rm Ni}}{\mu_{\rm Ni}}=0.52827,
\label{eq22} \\
&&
\frac{Z_{\rm Si}}{\mu_{\rm Si}}=0.50238,\qquad
\frac{N_{\rm Si}}{\mu_{\rm Si}}=0.50628.
\nonumber
\end{eqnarray}

The experimental data of Ref.~\cite{29} for $F_C^{\prime}(a)$
were taken at separation distances $a\geq 220\,$nm and found in
agreement with the Lifshitz theory \cite{23,39} generalized for
the case of magnetic materials. Specific characteristic feature
of the experiment with Au-Ni test bodies, as compared with the
case of two Au bodies considered in Sec.~III, is that here the
theoretical prediction over the entire measurement range does
not depend on whether the low-frequency behavior of the
dielectric permittivity of metals is described by the plasma
model or by the Drude model. This makes the theory-experiment
comparison fully transparent and independent of contribution
of any background effect (a discussion on the Drude and plasma
model approaches in the theory of dispersion forces can be
found in Ref.~\cite{22}).

Taking into account that in the limits of the total
experimental error,
$\Delta F_C^{\prime}(a)$, no additional force was observed, the
constraints on the parameters of the axion can  be again derived
from Eq.~(\ref{eq20}), where the gradient of the force due to
two-axion exchange is given by Eq.~(\ref{eq15}) with the
functions $X_s$ and $X_d$ defined in Eqs.~(\ref{eq16}) and
(\ref{eq21}), respectively.

The numerical analysis of the inequality (\ref{eq20})
shows that the strongest constraints are obtained at the
shortest experimental separation $a=220\,$nm, where the
total experimental error determined at a 67\% confidence
level is equal to $\Delta F_C^{\prime}=0.79\,\mu$N/m
\cite{29}.
This should be compared with the value of the Casimir
force gradient $F_C^{\prime}=131.03\,\mu$N/m
measured at this separation.
In Fig.~3, the solid lines show the constraints on the
pseudoscalar coupling constants $g_{ap(n)}^2/(4\pi)$,
as functions of the axion mass $m_a$,
which follow from the experiment \cite{29} with
Au-Ni test bodies.
The lines from bottom to top are plotted under the
conditions $g_{ap}^2=g_{an}^2$, $g_{an}^2\gg g_{ap}^2$
and $g_{ap}^2\gg g_{an}^2$, respectively.
As in Fig.~1, the regions above each line are prohibited by the
results of experiment \cite{29}, and the regions below each
line are allowed.

For comparison purposes, in the same figure we reproduce by the
dashed lines the constraints of Fig.~1 obtained in Sec.~III
under the same respective conditions from the experiment with
two Au surfaces. As can be seen in Fig.~3, the constraints
following from the experiment with Au-Ni test bodies \cite{29}
are up to a factor 1.5 weaker than the constraints from the
experiment with Au-Au bodies \cite{28,28a}.
This is mostly determined by the smaller density of Ni, as
compared with Au. At the same time, the constraints shown by
the solid and
dashed lines demonstrate the same behavior as functions
of $m_a$. Taking into account that the experiment with Au-Ni
test bodies is in a very good agreement with the Lifshitz
theory with no additional assumptions concerning the
low-frequency behavior of the dielectric permittivities,
this provides greater confidence in the constraints of
Sec.~III obtained from the experiment of Refs.~\cite{28,28a}
with two Au test bodies.

\section{Constraints from experiment with two nickel
test bodies}

In Refs.~\cite{30,31} the gradient of the Casimir force was
measured  between a Ni-coated hollow glass sphere of
$R=61.71\,\mu$m radius and $\Delta_s^{\!g}=5\,\mu$m
thickness and a Ni-coated
Si disc. As in the experiments of Refs.~\cite{28,28a,29},
the measurements were performed by means of dynamic AFM.
For technological purposes, there were, however, two
additional coatings on both test bodies in the experiment
of Refs.~\cite{30,31}. The outer layers of Ni were of
thicknesses $\Delta_s^{\!\rm Ni}=210\,$nm and
$\Delta_d^{\!\rm Ni}=250\,$nm on the spherical envelope and
on the disc, respectively. The thicknesses of Al-coatings
on both test bodies were equal
$\Delta_s^{\!\rm Al}=\Delta_d^{\!\rm Al}=40\,$nm.
There were also additional thin layers of Cr below Al layers
with thicknesses
$\Delta_s^{\!\rm Cr}=\Delta_d^{\!\rm Cr}=10\,$nm.
This layer structure should be taken into account in the
calculation of the additional force due to two-axion
exchange.

After application of Eq.~(\ref{eq13}) to each pair of material
layers forming the hollow sphere and the disc, one obtains
Eq.~(\ref{eq15}) for the gradient of the additional force.
In this case, however, due to a more complicated layer
structure, the functions $X_d$ and $X_s$ take the form
\begin{eqnarray}
&&
X_d(z)= C_{\rm Ni}\left(1-e^{-2z\Delta_d^{\!\rm Ni}}\right)
+C_{\rm Al}e^{-2z\Delta_d^{\!\rm Ni}}\left(1-e^{-2z\Delta_d^{\!\rm Al}}\right)
\nonumber \\
&&~~~
+C_{\rm Cr}e^{-2z(\Delta_d^{\!\rm Ni}+\Delta_d^{\!\rm Al})}
\left(1-e^{-2z\Delta_d^{\!\rm Cr}}\right)
+C_{\rm Si}
e^{-2z(\Delta_d^{\!\rm Ni}+\Delta_d^{\!\rm Al}+
\Delta_d^{\!\rm Cr})},
\nonumber \\[-1mm]
&&
\label{eq16b} \\[-2mm]
&&
X_s(z)= C_{\rm Ni}\left[\Phi(R,z)-e^{-2z\Delta_s^{\!\rm Ni}}
\Phi(R-\Delta_s^{\!\rm Ni},z)\right]
\nonumber \\
&&~~~
+C_{\rm Al}e^{-2z\Delta_s^{\!\rm Ni}}
\left[\Phi(R-\Delta_s^{\!\rm Ni},z)-e^{-2z\Delta_s^{\!\rm Al}}
\Phi(R-\Delta_s^{\!\rm Ni}-\Delta_s^{\!\rm Al},z)\right]
\nonumber \\
&&~~~
+C_{\rm Cr}e^{-2z(\Delta_s^{\!\rm Ni}+\Delta_s^{\!\rm Al})}
\left[\Phi(R-\Delta_s^{\!\rm Ni}-\Delta_s^{\!\rm Al},z)-
e^{-2z\Delta_s^{\!\rm Cr}}
\Phi(R-\Delta_s^{\!\rm Ni}-\Delta_s^{\!\rm Al}
-\Delta_s^{\!\rm Cr},z)\right]
\nonumber \\
&&~~~
+C_{g}e^{-2z(\Delta_s^{\!\rm Ni}+\Delta_s^{\!\rm Al}+\Delta_s^{\!\rm Cr})}
\left[
\vphantom{e^{-2z\Delta_s^{\! g}}}
\Phi(R-\Delta_s^{\!\rm Ni}-\Delta_s^{\!\rm Al}-\Delta_s^{\!\rm Cr},z)
\right.
\nonumber \\
&&~~~~~~~~~~
\left.
-
e^{-2z\Delta_s^{\! g}}
\Phi(R-\Delta_s^{\!\rm Ni}-\Delta_s^{\!\rm Al}-\Delta_s^{\!\rm Cr}
-\Delta_s^{\! g},z)\right],
\nonumber
\end{eqnarray}
\noindent
where the function $\Phi(r,z)$ is defined  in Eq.~(\ref{eq12}).

The coefficient $C_{\rm Cr}$, which was not used above, is given
by Eq.~(\ref{eq4}), where the values of $Z/\mu$ and $N/\mu$
for Cr can be found in Ref.~\cite{37},
\begin{equation}
\frac{Z_{\rm Cr}}{\mu_{\rm Cr}}=0.46518,\qquad
\frac{N_{\rm Cr}}{\mu_{\rm Cr}}=0.54379,
\label{eq24}
\end{equation}
\noindent
and the density of Cr is $\rho_{\rm Cr}=7.14\,\mbox{g/cm}^3$.

Measurements of the gradient of the Casimir force $F_C^{\prime}$
in Refs.~\cite{30,31} were performed at separations $a\geq 223\,$nm.
The experimental data were found in agreement with theoretical
predictions of the Lifshitz theory \cite{23,39}, generalized for
the case of magnetic materials, under the condition that the
low-frequency behavior of the dielectric permittivity of Ni is
described by the plasma model \cite{30,31}. The predictions of
the Lifshitz theory using the low-frequency behavior described by
the Drude model was excluded by the data. The same result
concerning the plasma and Drude models was obtained from the
experiment using two Au test bodies discussed in Sec.~III and
from three independent measurements of Refs.~\cite{40,41,42,43}
performed with two Au test bodies using an alternative
experimental technique (a micromachined oscillator).
The important characteristic feature of the case of two
magnetic surfaces is that here
$F_{C,D}^{\prime}>F_{C,p}^{\prime}$, where the indices
$D$ and $p$ denote which model, the Drude or the plasma,
is used in computations for the dielectric permittivity at
low frequencies. This is just the opposite to the case of two
nonmagnetic surfaces, Au for instance, where
$F_{C,D}^{\prime}<F_{C,p}^{\prime}$ \cite{30,31}.
Taking into account that the experiments of Refs.~\cite{28,28a}
with two Au surfaces, on the one hand, and of Refs.~\cite{30,31}
with two Ni surfaces, on the other hand, were performed using one
and the same setup (the dynamic AFM) and in both cases the
predictions of the Lifshitz theory $F_{C,p}^{\prime}$ was
confirmed, one can reliably exclude the role of any unaccounted
systematic effect in the theory-experiment comparison.
Note that the experiment \cite{56a}, claiming confirmation of the
Lifshitz theory combined with the Drude model, is not a direct
measurement of the Casimir force, but up to an order of magnitude
larger force of unknown origin. At separations below $3\,\mu$m
the results of this experiment are in fact uncertain because it
does not take into account glass imperfections, which unavoidably
present on lenses of large radii \cite{56b}. At larger separations
the measurement data of Ref.~\cite{56a} better agrees with the
plasma model \cite{56c}.

We also emphasize that the difference
$F_{C,p}^{\prime}-F_{C,D}^{\prime}$ cannot be approximated
either by the interaction due to two-axion exchange or by the
hypothetical Yukawa-type interaction \cite{24,25,26,27aa,27,28aa}.
Moreover, even if any other compensating interaction due to
some unknown agent existed and, being added to $F_{C,D}^{\prime}$,
brought the measurement data in
agreement with $F_{C,p}^{\prime}$ for Au-Au test bodies,
it would only increase the deviations between experiment and
theory for Ni-Ni test bodies and vice versa.
The reason is that for Au-Au bodies this difference corresponds
to the attractive force and for Ni-Ni to a repulsive.
Thus, the disagreement of the data with $F_{C,D}^{\prime}$
cannot be attributed to the influence of any additional force.

On this basis, we can obtain the constraints on the parameters
of an axion from Eq.~(\ref{eq20}) expressing the measure of
agreement of the measured data with $F_{C,p}^{\prime}$.
The strongest constraints follow at the
shortest experimental separation $a=223\,$nm, where the
total experimental error determined at a 67\% confidence
level is equal to $\Delta F_C^{\prime}=1.2\,\mu$N/m
\cite{30,31}.
This should be compared with the gradient of the Casimir
force measured at the shortest separation
$F_C^{\prime}=131.03\,\mu$N/m.
In Fig.~4 we show the obtained constraints on the
 coupling constants $g_{ap(n)}^2/(4\pi)$
by the solid lines,
as  functions of the axion mass $m_a$.
The lines from bottom to top are plotted under the
conditions $g_{ap}^2=g_{an}^2$, $g_{an}^2\gg g_{ap}^2$
and $g_{ap}^2\gg g_{an}^2$, respectively.
The regions above each line are prohibited and
below each line are allowed by the
results of experiment of Refs.~\cite{30,31}.
As can be seen in Fig.~4, the two upper lines, calculated
under conditions $g_{an}^2\gg g_{ap}^2$
and $g_{ap}^2\gg g_{an}^2$, almost overlap. This is caused by the
fact that for Ni the difference of the quantities $Z/\mu$ and
$N/\mu$ is much smaller than for Au.

In Fig.~4, we also reproduce by the
dashed lines the constraints of Fig.~1 obtained
under the same relationships between the coupling constants
 from the experiment with two Au test bodies \cite{28,28a}.
As is seen in Fig.~4, the experiment with two Ni test bodies
leads to weaker constraints than the
experiment with Au-Au bodies.
The qualitative character of the dashed lines in Fig.~4 is,
however, the same as of the solid lines, and weaker
constraints obtained are explained
by the smaller density of Ni, as compared with Au.
As noted above, the experiment with two Ni surfaces provides
further evidence for the absence of unaccounted systematic
effects in measurements by means of dynamic AFM.
Thus, the strongest constraints on the coupling constants of
an axion to nucleons obtained using this setup (see Fig.~1
and Table~I) can be considered as the best laboratory limits
within the region of axion masses from about $10^{-4}\,$eV
to 1\,eV.

\section{Conclusions and discussion}

In this paper, we have derived the constraints on the pseudoscalar
coupling constants of axion-like particles
to a proton and a neutron which
follow from measurements of the gradient of the Casimir force
in a sphere-plane geometry by means of dynamic AFM. For this
purpose, three recent experiments have been analyzed:
with an Au-coated sphere and an Au-coated disc \cite{28,28a},
with an Au-coated sphere and a Ni-coated disc \cite{29},
and with two Ni-coated test bodies \cite{30,31}.
Taking into consideration that the test bodies in measurements
of the Casimir interaction are unpolarized, we have calculated
the gradient of the additional force arising between them due
to two-axion exchange. This was done taking into account the layer
structure of both test bodies specific for each experiment.

In all three experiments considered the measurement data for
the gradient of the Casimir force were found in good agreement
with theoretical predictions of the Lifshitz theory of
dispersion forces, and no additional interaction was
observed. This allowed to constrain the gradient of the
additional force by the total experimental error in
measurements of the gradient of the Casimir force.
It was shown that the strongest constraints on the parameters
of an axion follow from the experiment with two Au test
bodies. However, two other experiments using magnetic (Ni)
surfaces provide further support to these constraints
by confirming the absence of any unaccounted systematic
effect in the measurements. The obtained constraints are
relevant to the wide region of axion masses from
$3\times 10^{-5}$ to 1\,eV and are determined at the 67\%
confidence level.
Using specific axion models, they can be conversed into limits
on the angle $\theta_{\rm eff}$ which measures {\it CP}
violating effects. For this purpose, however, one would need
to use respective constraints on a product of the pseudoscalar
and scalar axion-nucleon coupling constants \cite{48A}.

We have compared the derived constraints with the previously
known laboratory constraints of Ref.~\cite{20} following
from the measurements of thermal Casimir-Polder force \cite{21}.
It was found that the constraints obtained in
 this paper are stronger
over the entire range of the axion mass. Under the assumption
that $g_{an}=g_{ap}$, the largest strengthening by the factor
of 170 is achieved for the axion mass $m_a=0.3\,$eV.
Note that many constraints on axions with $m_a\lesssim 1\,$eV
using interactions with photons, electrons and nucleons
are based on the data of CERN axion solar telescope \cite{13},
globular cluster stars \cite{10,14}, and the duration of neutrino
burst of supernova 1987A \cite{11}. These constraints are
obtained from astrophysics and, broadly speaking,
may be more model-dependent \cite{47,48},
as compared to table-top laboratory experiments on
measuring the Casimir interaction.
In this respect the pure-laboratory experiment which searches for
photon oscillations into weakly interacting sub-eV particles is
of much interest \cite{51}.
It is important also that
our constraints are most strong in the
region of axion masses
$4\times 10^{-5}\,\mbox{eV}<m_a<0.3\,$eV
partially overlapping with
the axion window \cite{10} where model-independent constraints on the axion
parameters are rather weak.

One can conclude that laboratory experiments on
measuring the Casimir interaction have considerable opportunity in
the search for an axion and further constraining its parameters.

\section*{Acknowledgments}

The authors of this work acknowledge CNPq (Brazil) for
 partial financial support.
G.L.K.\ and V.M.M.\ are grateful to U.~Mohideen for
useful discussions of his experiments.
They also acknowledge the Department
of Physics of the Federal University of
Para\'{\i}ba (Jo\~{a}o Pessoa, Brazil) for hospitality.


\begin{figure}[b]
\vspace*{-8cm}
\centerline{\hspace*{2cm}
\includegraphics{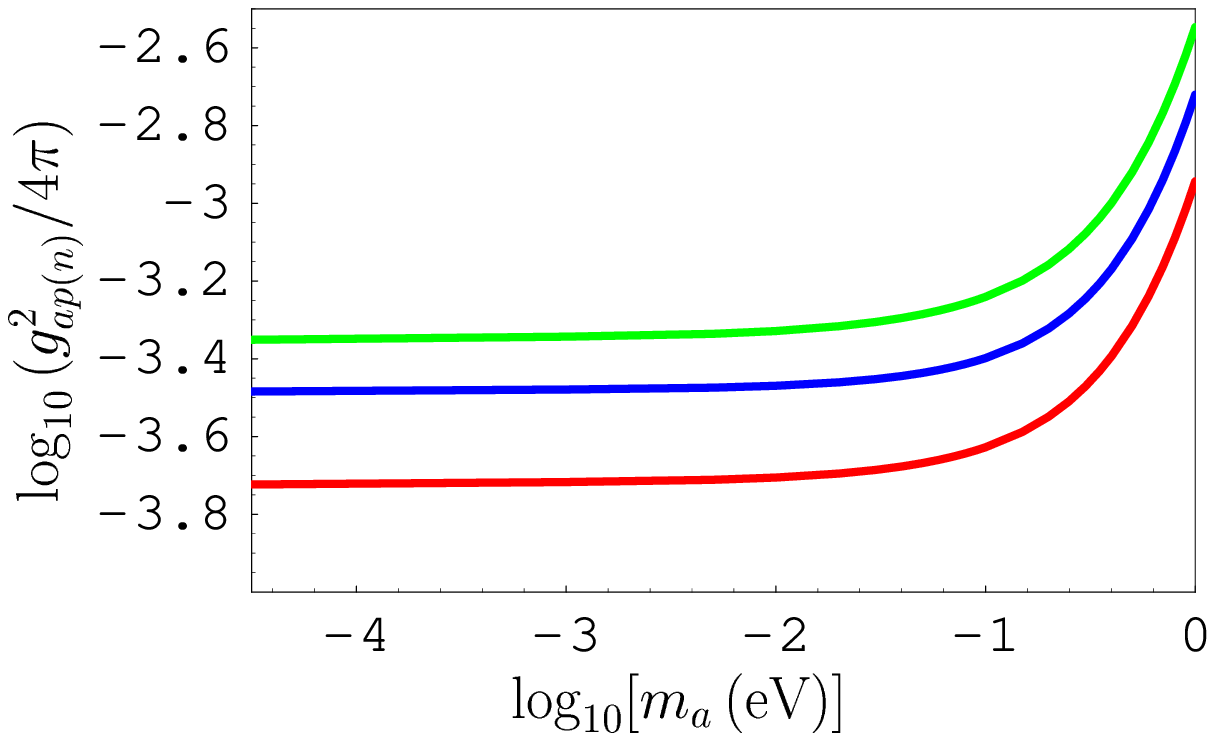}
}
\vspace*{-9cm}
\caption{(Color online)
Constraints on the  coupling constants
of an axion to a proton or a neutron obtained from
measurements of the gradient of the Casimir force
between Au surfaces
as functions of the axion mass.
The lines from bottom to top are plotted under the
conditions $g_{ap}^2=g_{an}^2$, $g_{an}^2\gg g_{ap}^2$,
and $g_{ap}^2\gg g_{an}^2$, respectively.
 The regions  above each line
are prohibited and below each line are allowed.
}
\end{figure}
\begin{figure}[b]
\vspace*{-8cm}
\centerline{\hspace*{2cm}
\includegraphics{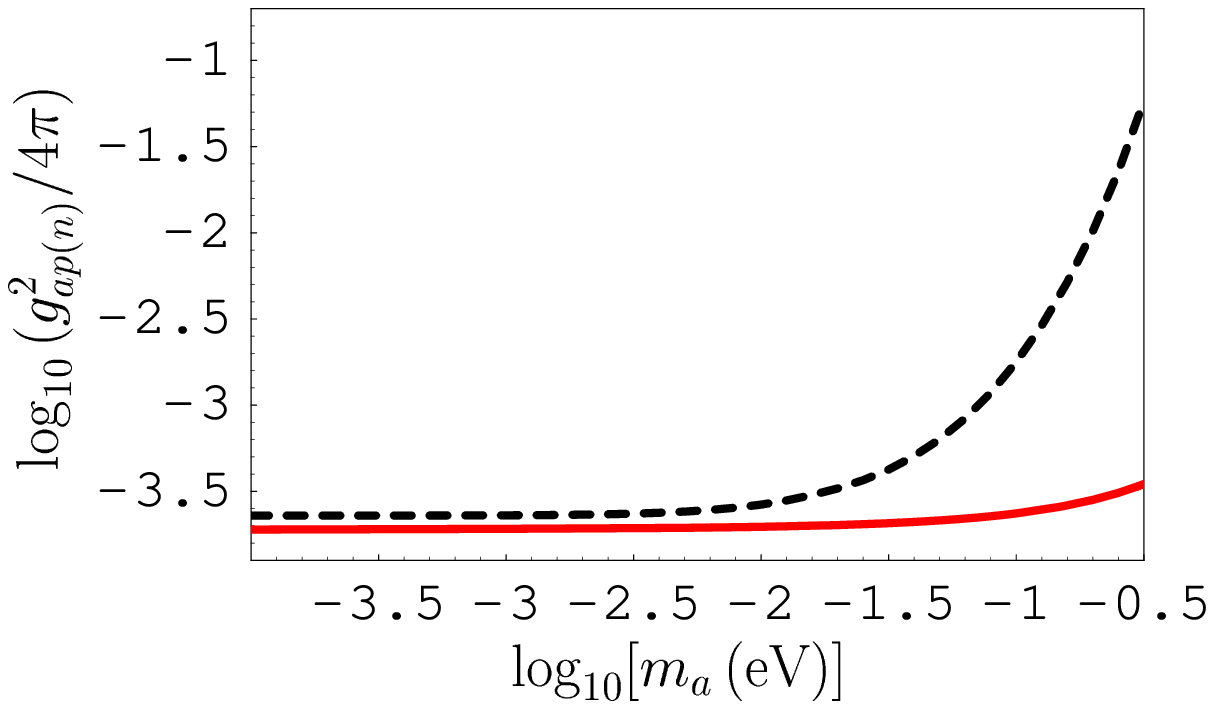}
}
\vspace*{-9cm}
\caption{(Color online)
Constraints on the coupling constants
of an axion to a nucleon under the
condition $g_{ap}^2=g_{an}^2$
as functions of the axion mass.
The solid and dashed lines show the results obtained
in this paper from measurements of the gradient of
the Casimir force between Au surfaces \cite{28,28a} and
in Ref.~\cite{20} from
measurements of the thermal Casimir-Polder force
\cite{21}, respectively.
 The regions of the plane above each line
are prohibited and below each line are allowed.
}
\end{figure}
\begin{figure}[b]
\vspace*{-8cm}
\centerline{\hspace*{2cm}
\includegraphics{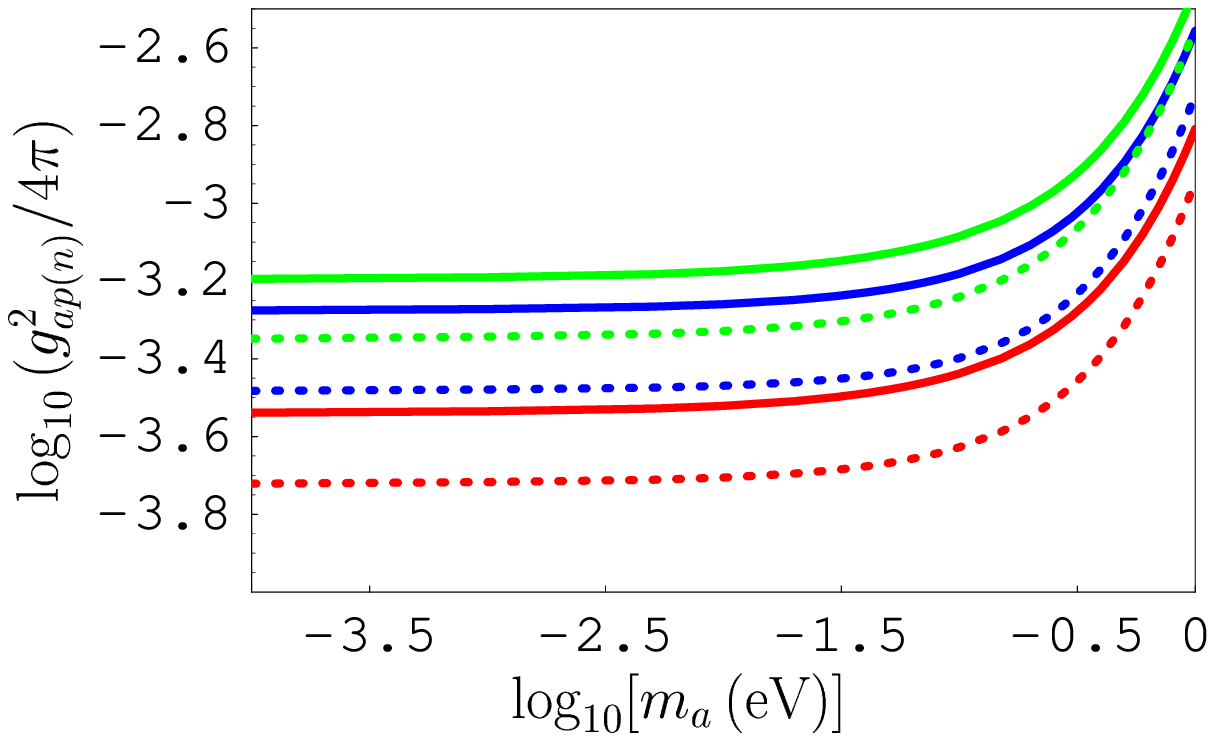}
}
\vspace*{-9cm}
\caption{(Color online)
Constraints on the  coupling constants
of an axion to a proton or a neutron following from
the experiment with Au-Ni test bodies (solid lines)
and Au-Au test bodies (dashed lines) are shown
as functions of the axion mass.
The lines from bottom to top are plotted under the
conditions $g_{ap}^2=g_{an}^2$, $g_{an}^2\gg g_{ap}^2$,
and $g_{ap}^2\gg g_{an}^2$, respectively.
 The regions above each line
are prohibited and below each line are allowed.
}
\end{figure}
\begin{figure}[b]
\vspace*{-8cm}
\centerline{\hspace*{2cm}
\includegraphics{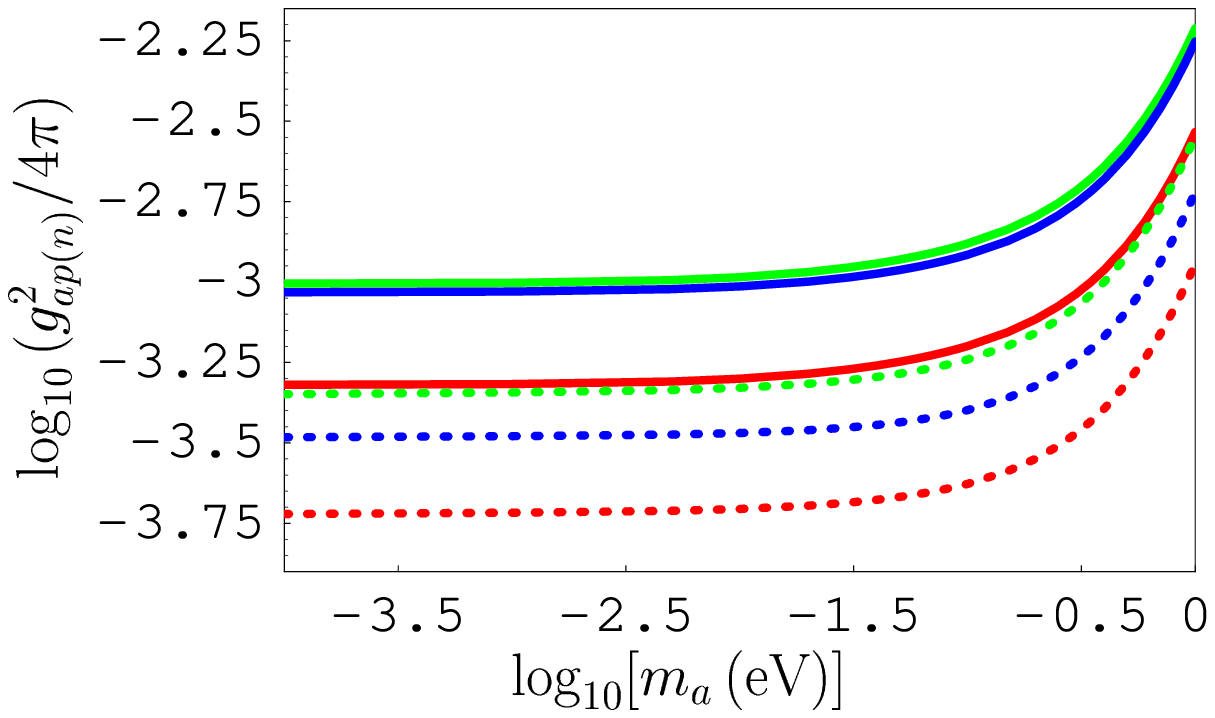}
}
\vspace*{-9cm}
\caption{(Color online)
Constraints on the  coupling constants
of an axion to a proton or a neutron following from
the experiment with Ni-Ni test bodies (solid lines)
and Au-Au test bodies (dashed lines) are shown
as functions of the axion mass.
The lines from bottom to top are plotted under the
conditions $g_{ap}^2=g_{an}^2$, $g_{an}^2\gg g_{ap}^2$,
and $g_{ap}^2\gg g_{an}^2$, respectively.
 The regions above each line
are prohibited and below each line are allowed.
}
\end{figure}
\begingroup
\squeezetable
\begin{table}
\vspace*{2cm}
\caption{Maximum values of the coupling constants of
an axion to a proton and a neutron, allowed by
measurements of the gradient of the Casimir force
between Au surfaces, are calculated for different axion
masses (column 1) under the conditions
$g_{ap}^2=g_{an}^2$ (column 2),
$g_{an}^2\gg g_{ap}^2$ (column 3),
and $g_{ap}^2\gg g_{an}^2$ (column 4).
}
\begin{ruledtabular}
\begin{tabular}{cccc}
$m_a\,$(eV) & $\frac{g_{ap}^2}{4\pi}=\frac{g_{an}^2}{4\pi}$ &
$\frac{g_{an}^2}{4\pi}\gg\frac{g_{ap}^2}{4\pi}$ &
$\frac{g_{ap}^2}{4\pi}\gg\frac{g_{an}^2}{4\pi}$ \\[1mm]
\cline{1-4}
$\leq0.001$& $1.92\times 10^{-4}$&$3.32\times 10^{-4}$&
$4.54\times 10^{-4}$ \\
0.01 &$1.97\times 10^{-4}$&$3.39\times 10^{-4}$&
$4.69\times 10^{-4}$ \\
0.05 &$2.15\times 10^{-4}$&$3.66\times 10^{-4}$&
$5.18\times 10^{-4}$ \\
0.1 &$2.36\times 10^{-4}$&$4.00\times 10^{-4}$&
$5.74\times 10^{-4}$ \\
0.2&$2.83\times 10^{-4}$&$4.76\times 10^{-4}$&
$6.95\times 10^{-4}$ \\
0.3&$3.38\times 10^{-4}$&$5.69\times 10^{-4}$&
$8.36\times 10^{-4}$ \\
0.4&$4.05\times 10^{-4}$&$6.79\times 10^{-4}$&
$1.00\times 10^{-3} $\\
0.5&$4.84\times 10^{-4}$&$8.10\times 10^{-4}$&
$1.20\times 10^{-3} $\\
0.6&$5.77\times 10^{-4}$&$9.65\times 10^{-4}$&
$1.43\times 10^{-3} $\\
0.7&$6.86\times 10^{-4}$&$1.15\times 10^{-3}$&
$1.71\times 10^{-3} $\\
0.8&$8.15\times 10^{-4}$&$1.36\times 10^{-3}$&
$2.03\times 10^{-3} $\\
0.9&$9.65\times 10^{-4}$&$1.61\times 10^{-3}$&
$2.40\times 10^{-3} $\\
1.0&$1.14\times 10^{-3}$&$1.90\times 10^{-3}$&
$2.84\times 10^{-3}$
\end{tabular}
\end{ruledtabular}
\end{table}
\endgroup

\end{document}